\documentclass[twocolumn,prl,showpacs,preprintnumbers,superscriptaddress]{revtex4-1}

\usepackage{graphicx}
\usepackage{amsmath} 
\usepackage{amssymb}
\usepackage{amsfonts}
\usepackage{mathrsfs}
\usepackage{units}
\usepackage{multirow}

\usepackage{dcolumn}
\usepackage{bm}
\usepackage{rotating} 

\newcommand{\disregard}[1]{}

\begin{document}

\title{Beta-decay study within multi-reference density functional theory and beyond
}

\author{M. Konieczka}
\affiliation{Faculty of Physics, University of Warsaw, ul. Pasteura 5,
PL-02-093 Warsaw, Poland}

\author{P. B{\c a}czyk}
\affiliation{Faculty of Physics, University of Warsaw, ul. Pasteura 5,
PL-02-093 Warsaw, Poland}

\author{W. Satu{\l}a}
\affiliation{Faculty of Physics, University of Warsaw, ul. Pasteura 5,
PL-02-093 Warsaw, Poland}
\affiliation{Helsinki Institute of Physics, P.O. Box 64, FI-00014 University of Helsinki, Finland}

\date{\today}

\begin{abstract}
Pioneering study of Gamow-Teller (GT) and Fermi matrix elements (MEs) using no-core-configuration-interaction formalism
rooted in multi-reference density functional theory is presented. After successful test performed for
$^6$He$\rightarrow$$^6$Li $\beta$-decay, the model is applied to compute MEs in the $sd$- and $pf$-shell
$T$=1/2 mirror nuclei. The calculated GT MEs and the isospin-symmetry-breaking corrections
to the Fermi branch are found to be in a very good agreement with shell-model predictions in spite of fundamental
differences between these models concerning model space, treatment of correlations or inclusion of a core. This result
indirectly supports the two-body current based scenarios behind the quenching of axial-vector coupling
constant.
\end{abstract}

\pacs{21.10.Hw, 
21.60.Jz, 
21.30.Fe, 
23.40.Hc, 
24.80.+y 
}
\maketitle

The atomic nuclei are unique laboratories to study fundamental processes and search for possible signals of
{\it new physics\/} beyond the Standard Model in ways that are complementary or even superior to other sciences.
This is due to enhanced sensitivity of specific isotopes to fundamental symmetries caused, in particular, by
intrinsic-symmetry-related or purely accidental (near-)degeneracies of nuclear states.
For example, parity doublets caused by stable octupole deformation increase sensitivity
to search for the violation of CP-symmetry which is responsible for matter over anti-matter dominated
Universe~\cite{(Dob05e)}. Accidental near-degeneracy of 3/2$^+$ and $5/2^+$ levels in $^{229}$Th,
which are separated only by 7.6$\pm$0.5\,eV, opens up a possibility for high-precision measurement of the temporal variation
of fine-structure constant with much higher sensitivity as compared to the atomic transitions~\cite{(Ber10)}.  Last but not
least, nuclear physics input is critical in an ongoing hunt for a weakly-interacting massive particle (WIMP),
a candidate for dark matter, in direct detection experiments measuring the recoil energy deposited when
WIMP is scattered off the nucleus, see~\cite{(Hof15)} and refs. quoted therein.

Traditionally, the atomic nuclei are used to study the weak interaction. A flagship
example is the superallowed $I$=0$^+$$\rightarrow$$I$=0$^+$ $\beta$-decay
among the members of the isobaric triplets $T$=1.
With small, of order of a percent, theoretical corrections accounting for
radiative processes and isospin symmetry breaking (ISB), these semileptonic pure Fermi (vector) decays
allow to verify the conserved vector current (CVC) hypothesis  with a very high precision. In turn, they
provide the most precise values of the strength of the weak force, $G_{\text{F}}$, and
of the leading element, $V_{\text{ud}}$, of the Cabbibo-Kobayashi-Maskawa (CKM) matrix, see~\cite{(Har14)}
for a recent review.

The $T$=1/2 mirror nuclei offer an alternative way to test the CVC hypothesis~\cite{(Nav09a)}.
These nuclei decay via the mixed Fermi and Gamow-Teller (GT) transitions. Hence,
apart from the radiative and the ISB theoretical corrections,
the final values of $G_{\text{F}}$ and $V_{\text{ud}}$ depend on the ratio of statistical rate functions
for the axial-vector and vector interactions, $f_{\text{A}}/f_{\text{V}}$, and the ratio of nuclear matrix elements
$\rho \approx \lambda M_{\text{GT}}/M_{\text{F}}$ where $\lambda =g_{\text{A}}/g_{\text{V}}$ denotes the ratio of axial-vector
and vector coupling constants.

The CVC hypothesis implies that the vector coupling constant is a true constant $g_{\text{V}}$=1. The axial-vector current
is partially conserved meaning that the coupling constant gets renormalized in nuclear medium. The
effective axial-vector coupling constant, $g^{\text{(eff)}}_A$=$q g_{\text{A}}$, is quenched by an $A$-dependent factor, $q$,
with respect to its free neutron decay value $g_{\text{A}} \approx -1.2701(25)$. Quenching factors deduced
from comparisons between the large-scale nuclear-shell-model (NSM) calculations and experiment are: $q\approx 0.82$,
$q\approx 0.77$~\cite{(Bro85)} and $q\approx 0.74$~\cite{(Mar96s)} in the $p$-, $sd$-, and $pf$-shell, respectively.
In heavier, $A$=100-134, nuclei, the average quenching is $q \approx 0.48$~\cite{(Pir15)}. This result is consistent, up to
a theoretical uncertainty, with the result of Ref.~\cite{(Cau12)}. Even stronger quenching, $g_{\text{A}}A^{-0.18}$,
is used in the IBM-2 model~\cite{(Bar15)}.

The question about physical causes
of the quenching has no unique answer. The quenching is usually related to: ({\it i\/}) missing correlations in the wave function,
({\it ii\/}) truncation of model space or ({\it iii\/}) to a very fashionable nowadays renormalization of the GT operator due to
the two-body currents~\cite{(Men11),(Eks14s)}. Scenarios involving non-nucleonic degrees of freedom like $NN\rightarrow N\Delta$
excitations are shown to contribute only rather weakly~\cite{(Ich06)}.

Proper understanding of the quenching is essential for many branches of modern physics from
modeling of astrophysical processes in stars to elusive neutrinoless double beta decay which depends
on the fourth power of $g^{\text{(eff)}}_A$ and is therefore particularly sensitive to $q$.
In order to address the quenching, it is of paramount importance to
investigate the GT matrix elements (MEs) using diverse theoretical models.
The goal of this work is to communicate the pioneering application of the multi-reference
density functional theory (MR-DFT) rooted no-core-configuration-interaction (NCCI) approach to study
$\beta$-decay, with a particular emphasis on the GT process. After a short introduction to the model
we shall present the numerical results starting with the $\beta$-decay of $^6$He in order to test
reliability of the model. Afterwards, the model will be applied to the  $sd$- and lower $pf$-shell $T$=1/2
mirror nuclei, where both the GT MEs and the Fermi MEs will be computed.

The NCCI models rooted in MR-DFT offer nowadays an interesting alternative
to the conventional nuclear shell model~\cite{(Bal14bs),(Sat14s),(Sat15)}. Firstly, they are capable of treating rigorously
both the fundamental (spherical, particle number) as well as approximate (isospin) symmetries. Secondly, by invoking the
generator coordinate method and/or mixing of discrete (quasi)particle-(quasi)hole (or particle-hole) configurations, they
allow to incorporate important correlations into the nuclear wave function. Thirdly, they can be applied
to any nucleus irrespectively of $A$ and the neutron and proton number parities. Moreover, by construction, they are able to capture
core-polarization effects resulting from a subtle interplay between the long-range and short-range nucleon-nucleon forces, what
is of critical importance for the calculation of isospin impurities and ISB corrections~\cite{(Sat11s)}.

The NCCI formalism developed by our group~\cite{(Sat14s),(Sat15)} involves the angular momentum and isospin projections
and subsequent mixing of states having good angular momentum and properly treated isospin.
It proceeds in three distinct steps. First, we compute self-consistently a set of $k$ Hartree-Fock (HF) (multi)particle-(multi)hole
configurations, $\{|\varphi_j\rangle \}_{j=1}^k$, relevant for the problem under study.
The Slater determinants $\{|\varphi_j\rangle \}_{j=1}^k$ are calculated
using true Skyrme interaction in order to avoid singularities at the next stage, at which we apply the angular momentum
and isospin projections to determine, for each $j$, the family of states $\{ |\varphi_j; IMK; TT_z\rangle \}$ having good
isospin $TT_z$, angular momentum $IM$, and angular-momentum projection on the intrinsic axis $K$. Subsequently,
the states  $\{ |\varphi_j; IMK; TT_z\rangle \}$ are mixed in order to account for the $K$ and isospin mixing.
This gives a set of good angular-momentum states $\{ |\varphi_j;\, IM; T_z\rangle^{(i)}\}_{i=1}^{l_j}$ for each
HF configuration $j$. The set is non-orthogonal and, in general, overcomplete. In the final step, the states
are mixed over different configurations by solving the Hill-Wheeler-Griffin (HWG) equation, $\hat H u=E N u$,
with the same Skyrme interaction that was used at the HF level.
The HWG equation is solved in the {\it collective space\/} spanned  by the {\it natural states\/} corresponding
to non-zero eigenvalues $n$ of their norm matrix $N$. The same technique is used in the code to handle the $K$-mixing
alone as described in detail in Ref.~\cite{(Dob09ds)}.

On exit, the  NCCI code provides eigenfunctions that are labeled by the index $n$ numbering eigenstates in ascending order
according to their energies and the strictly conserved quantum numbers
$I$, $M$, and $T_z=(N-Z)/2$. The eigenstates can be decomposed in
original projected (non-orthogonal) basis:
\begin{eqnarray}\label{KTmix}
|n; \, IM; \, T_z\rangle &=&  \sum_{i,j}
   a^{(n; IM; T_z)}_{ij} |\varphi_j;\, IM; T_z\rangle^{(i)}  \\
    &=&  \sum_{i,j} \sum_{K, T\geq |T_z|}
   f^{(n; IM; T_z)}_{ijKT} \hat P^T_{T_z T_z} \hat P^I_{MK} |\varphi_j \rangle \nonumber \, ,
\end{eqnarray}
where $\hat P^T_{T_z T_z}$ and $\hat P^I_{MK}$ stand for the isospin
and angular-momentum projection operators, respectively.
This form is particularly useful to compute MEs of the GT operator:
\begin{equation}\label{matel}
M_{\mu, \nu} \equiv \mp \langle n'; \, I'M'; \, T'_z | {\cal O}_{\mu,\nu} | n; \, IM; \, T_z\rangle .
\end{equation}
where ${\cal O}_{\mu,\nu} = \frac{1}{\sqrt{2}}\sum_k^A \hat\tau_{1\mu}^{(k)} \hat \sigma_{1\nu}^{(k)}$ is
expressed by means of one-body spherical tensors. The isospin index above is fixed $\mu =\pm 1$ and it
determines the overall phase factor. The matrix element (\ref{matel}) fulfills the
Wigner-Eckart theorem:
\begin{equation}\label{WiEc}
M_{\mu, \nu} =  \frac{1}{\sqrt{2I'+1}} \, C^{I'M'}_{IM,1\nu} \, \langle n', I' || {\cal O}_{\mu} || n, I \rangle
\end{equation}
where $C^{I'M'}_{IM,1\nu}$ stands for the Clebsch-Gordan coefficient. The reduced matrix element
equals:
\begin{eqnarray} \displaystyle
  &\,& \langle n',  I' || {\cal O}_{\mu}   || n, I \rangle =
 \mp \sum_{ijKT} \sum_{i'j'K'T'}
f^{(n'; I'M'; T_z')^{\,*}}_{i'j'K'T} f^{(n; IM; T_z)}_{ijKT}  \nonumber \\
&\,& \sqrt{2I'+1} \, C^{T'T_z'}_{TT_z,1\mu} \sum_{\eta,\xi} C^{T'T_z'}_{TT_z'-\eta,1\eta}
C^{I'K'}_{IK'-\xi,1\xi} J_{\eta, \xi; j,j' }^{(TT_zT_z';IKK')}.
\end{eqnarray}
The integral, $J$, runs over the beta Euler angle in isospace, $\beta_T$, and the Euler
angles in space  $\Omega = (\alpha, \beta, \gamma)$:
\begin{eqnarray}\label{integ}
J_{\eta,\xi; j,j'}^{(TT_zT_z';IKK')}
 =  &\,& \frac{2T+1}{2} \int_0^\pi d\beta_T \sin\beta_T \, d^T_{T_z'-\eta, T_z} \nonumber \\
\frac{2I+1}{8\pi^2} &\,& \int d\Omega \, D^{I^{\,*}}_{K'-\xi, K} \langle \varphi_{j'} | {\cal O}_{\eta,\xi}  | \tilde{\tilde{\varphi}}_{j} \rangle.
\end{eqnarray}
where $d^T_{T_z',T_z}$ and $D^{I}_{M, K}$ are the Wigner functions and
${\tilde {\tilde \varphi}}_{j}$ denotes Slater determinant rotated in space
and isospace. Mean-field (MF) matrix elements in Eq.~(\ref{integ}) can be expressed by means of transition densities. The formulas are somewhat lengthy
and will be published in our forthcoming paper. All integrals appearing above
can be calculated exactly by applying appropriate quadratures, see~\cite{(Dob09ds)} for further details.

The Wigner-Eckart relation (\ref{WiEc}) implies that the total probability of decay summed up over the components $\nu$ of the
operator ${\cal O}_{\mu,\nu}$ and over polarizations of the final state $M'$ is:
\begin{eqnarray}\label{BGT}
B({\cal O}_\mu; &\,& n,I \rightarrow n',I') \nonumber \\
&\,& = g_{\text{A}}^2 \frac{ |\langle n', I' || {\cal O}_{\mu} || n, I \rangle|^2}{2I+1} \equiv g_{\text{A}}^2 \frac{ |M_{\text{GT}}|^2}{2I+1}.
\end{eqnarray}

The calculations discussed below were done using a new unpublished version of the HFODD solver~\cite{(Dob09ds),(Sch12s)},
which was equipped with the NCCI module. In order to track mean-field (MF) configurations and facilitate convergence properties
at the MF level all reference states used in the NCCI calculations has been self-consistently calculated assuming parity and
signature symmetries. In the calculations we have used a basis consisting either $N$=10 or 12 spherical harmonic
oscillator (HO) shells. The calculations has been performed using either the SV Skyrme force of Ref.~\cite{(Bei75)} or a variant of this force,
dubbed SV$_{\text{SO}}$, having 20\% stronger spin-orbit interaction. The latter force was introduced to
improve slightly the single-particle (s.p.) properties of the SV, in particular, by shifting the $d_{3/2}$ subshell with respect to the $s_{1/2}$ subshell. The near-degeneracy of these subshells in the SV energy density functional (EDF) causes strong  mixing and,
in turn, leads to unphysically large ISB corrections in the superallowed $0^+$$\rightarrow$$0^+$ $\beta$-decay in $A$=38
isospin triplet~\cite{(Sat11s)}. The SV$_{\text{SO}}$ force does not cure the problem of incorrectly placed s.p. levels but makes the
spectrum slightly more realistic. Indeed, in $^{40}$Ca, the $s_{1/2}$ and $d_{3/2}$ neutron subshells are separated by
0.22MeV in the SV EDF and 1.18MeV in the SV$_{\text{SO}}$ EDF, respectively. For comparison, the experimental splitting
is 2.55MeV~\cite{(Oro96)}.

\begin{figure}[thb]
\centering
\includegraphics[width=0.8\columnwidth]{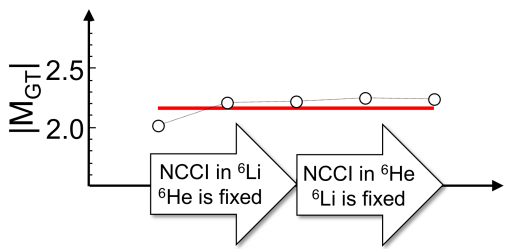}
\caption{(Color online) GT ME for beta decay of
$^6$He in function of a number of configurations taken in the NCCI calculations in $^6$Li and
$^6$He. Solid line shows the experimental value of Ref.~\protect{\cite{(Kne12s)}}.
}
\label{fig01}
\end{figure}

Recently, Knecht {\it et al.}~\cite{(Kne12s)} performed high-precision measurement of the $0^+$$\rightarrow$$1^+$ beta decay of $^6$He, which
proceeds exclusively to the ground state (GS) of $^6$Li and determined the corresponding GT matrix element
$|M_{\text{GT}}|$=2.1645(43) assuming $g_{\text{A}}$=-1.2701(25). This is an excellent test case for our model mainly because of
a limited number of $ph$ configurations that can contribute in these $p$-shell nuclei. The results of the NCCI calculations
performed for this transition are depicted in Fig.~\ref{fig01}. The figure shows the calculated GT ME versus a number
of configurations taken in the mixing. The very left point corresponds to a situation, where no mixing was performed in neither of the
nuclei. In this limit, called hereafter the MR-DFT limit, the HF reference states were selected based ultimately on the energy
criterion. Note, that already in this limit the calculated ME is in fair agreement with the empirical value underestimating
it by $\approx$7\%. Next, keeping the  wave function of $^6$He fixed, we have attempted to correlate
the wave function of $^6$Li by admixing $1^+$ states projected from the lowest $ph$
configuration (second point) and  from the first two lowest $ph$ configurations (third point).
This caused an increase of the ME to 2.208 and 2.223, respectively, i.e. circa 3\% above the experiment,
see Fig.~\ref{fig01}. At this point we freeze the wave function of $^6$Li
and attempt to correlate the wave function of $^6$He (last two points). This weakly influences the ME
giving eventually 2.238.
The test shows that the model is capable to capture main features of the wave functions that are important for reliable
reproduction of the GT ME and provides stable predictions in function of a number of admixed configurations.

\begin{figure}[htb]
\centering
\includegraphics[width=0.8\columnwidth]{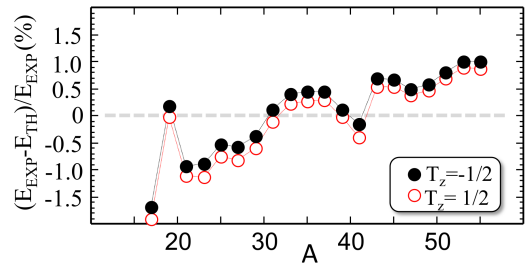}
\caption{(Color online) Theoretical binding energies of $T$=1/2 mirror nuclei
calculated using the NCCI framework. The results are shown relative to the
experimental data.
}
\label{fig02}
\end{figure}

Encouraged by the result obtained for the $^6$He decay,
we have performed systematic study of the GT and Fermi MEs
for the GS$\rightarrow$GS transitions in the $T$=1/2 mirror nuclei covering the $sd$- and lower $pf$-shell nuclei
from $A=17$ till $55$. All results shown below were obtained using the SV$_{\text{SO}}$ EDF. This functional,
apart from slightly more realistic s.p. levels, is also superior in reproducing binding energies (BE) in comparison to the SV EDF.
The ability to reproduce masses is considered to be one of the most important signatures of the quality of DFT-rooted models.
The calculated BE relative to empirical results are depicted in Fig.~\ref{fig02}. Although the theory tends to overbind the lightest
species and underbind the heavier, the overall agreement is at a quite impressive level of $\pm 1$\%. It is better almost
by a factor of two than the level of agreement obtained for the SV interaction.

\begin{figure}[htb]
\centering
\includegraphics[width=0.8\columnwidth]{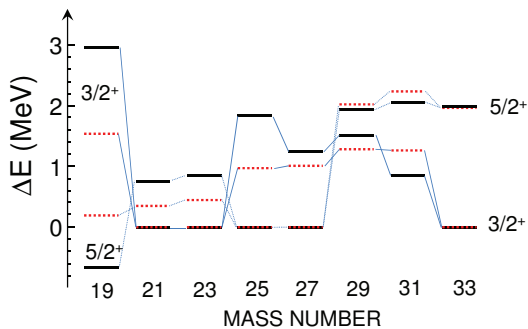}
\caption{(Color online)  Excitation energies of the lowest $3/2^+$ and $5/2^+$ states in $sd$-shell $T$=1/2, $T_z$=1/2
nuclei ranging from $A$=19 till 33. The calculated (experimental) levels are marked by thick solid (dashed) lines,
respectively. The calculated states come from MR-DFT.
Theoretical energies have been normalized to the experimental ground states.
}
\label{fig03}
\end{figure}

It appears also that the SV$_{\text{SO}}$ has reasonable spectroscopic properties.
The theory is  able to reproduce the GS spins already at the MR-DFT level with the
exception of $A=19$ case, where the model predicts $I=5/2^+$ instead of $I=1/2^+$ to be
the GS spin. The energy spectra are, in general,  in fair agreement with data. For the sake of
illustration Fig.~\ref{fig03} shows theoretical and experimental $I$=3/2$^+$ and 5/2$^+$ states in the
lower $sd$-shell nuclei. Similar agreement is obtained for heavier nuclei.

\begin{figure}[htb]
\centering
\includegraphics[width=0.75\columnwidth]{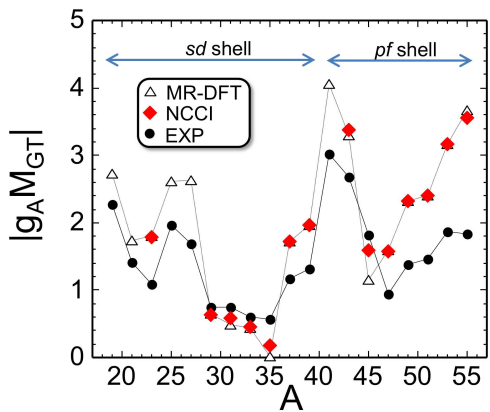}
\caption{(Color online) Gamow-Teller matrix elements calculated using the MR-DFT (triangles) and NCCI (diamonds) approaches
in comparison with experimental data (dots) taken from Ref.~\cite{(Bro85)} ($sd$-shell) and Refs.~\cite{(Mar96s),(Sek87)} ($pf$-shell).
}
\label{fig04}
\end{figure}

Fig.~\ref{fig04} shows the GT MEs, $|g_{\text{A}} M_{\text{GT}}|$, calculated using the MR-DFT and
the NCCI models. The NCCI calculations involve typically four-five low-lying
$ph$ MF configurations. The results of both approaches are strikingly similar except for the
$^{45}$V$\rightarrow$$^{45}$Ti transition. Both models systematically overestimate the data beside
the nuclei ranging from $A$=29 to 35. Strong suppression of the GT MEs in this mass range is related to 
aforementioned clustering of the $s_{1/2}$ and $d_{3/2}$ subshells in MF calculations. Proximity of these two subshells causes strong
mixing, which has destructive impact on the calculated MEs. Comparison of the MR-DFT results obtained
using the SV and SV$_{\text{SO}}$ EDFs supports this conclusion. Indeed, the MEs calculated using these two
functionals are almost identical everywhere except for the mass region discussed above.

\begin{figure}[htb]
\centering
\includegraphics[width=0.75\columnwidth]{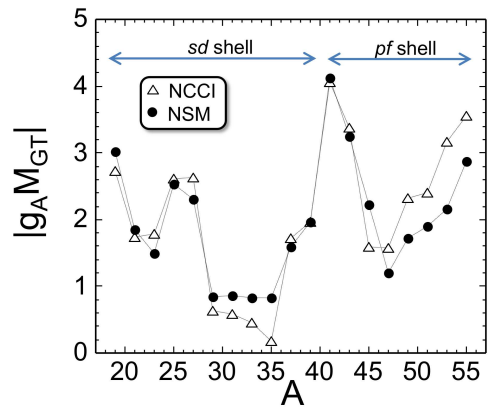}
\caption{
GT MEs calculated using the NCCI model (triangles) in comparison to the NSM results of
Ref.~\cite{(Bro85)} ($sd$-shell) and Refs.~\cite{{(Mar96s)},{(Sek87)}} ($pf$-shell).
}
\label{fig05}
\end{figure}

Fig.~\ref{fig05} shows the NCCI results in comparison to the NSM calculations of Ref.~\cite{(Bro85)} ($sd$-shell)
and Refs.~\cite{(Mar96s),(Sek87)} ($pf$-shell). The two sets of calculations are very consistent with each other.
Indeed, linear fit to the MR-DFT (NCCI) GT MEs gives $q$=0.77 (0.78) in the $sd$-shell (excluding problematic $A=31-35$ cases)
and $q$=0.75 (0.69) in the lower $pf$-shell, respectively. These values agree almost perfectly with the NSM quenching
in spite of numerous differences between the models.  In particular, our model: ({\it i\/}) includes the core and
takes into account core polarization effects, ({\it ii\/}) accounts for correlations in a different, more schematic way,
than the shell-model, ({\it iii\/}) uses functionals, which were not optimized for the NCCI calculations or
({\it iv\/}) uses completely different model space. In order to address the latter point, we have performed
additional set of MR-DFT calculations using larger basis, consisting $N$=12 spherical HO shells. We found that the increase of the
basis size has almost no impact on the calculated MEs. Note also, a systematic difference between the shell-model
and NCCI results in the heaviest calculated nuclei. The origin of this difference requires deeper study.

\begin{figure}[htb]
\centering
\includegraphics[width=0.75\columnwidth]{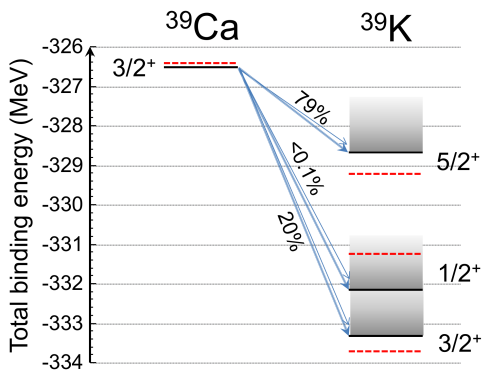}
\caption{(Color online) Ikeda sum rule for the GS of $^{39}$Ca. Thick horizontal lines
indicate the theoretical (solid) and experimental (dashed) total BE
of the GS in $^{39}$Ca and the lowest $1/2^+$, $3/2^+$, and $5/2^+$ states in $^{39}$K.
The numbers over the arrows indicate the total calculated GT strength for each $I$.
Multiple arrows and shadowing indicate that the strength is distributed over several states.
}
\label{fig06}
\end{figure}

The Ikeda sum rule is an important indicator of the quality of theoretical models.
For the $T$=1/2 mirrors it takes particularly simple form: $\sum_{n',I'} B({\cal O}_+; n,I \rightarrow n',I') = 3$.
Systematic study of the Ikeda sum rule with the present formalism involving the isospin and angular momentum projections is
CPU expensive. Hence, it was limited here to one of the simplest cases of $A$=39 nuclei.
In this case, inclusion of all possible $ph$ excitations within the $sd$-shell exhausts 99\% of the sum rule as
illustrated in Fig.~\ref{fig06}. It is worth mentioning that inclusion of $ph$ excitations
between the spin-orbit partners $d_{5/2}$$\rightarrow$$d_{3/2}$ is crucial for the sum rule. More systematic study of the sum
rules will be done with the variant involving only the angular momentum projection.

The use of NCCI approach involving both the isospin and angular momentum projected states
is absolutely necessary to study Fermi transitions and, in particular, to extract the ISB
corrections to the Fermi branch of $\beta$-decay. Let us recall that the ISB corrections are needed to study
the CVC hypothesis and the CKM matrix via the transitions
in the mirrors, see Ref.~\cite{(Nav09a)}. The results obtained in this study are collected in Tab.~\ref{tab01}.
It is beneficial to see that our corrections are very consistent with the NSM results of Ref.~\cite{(Sev08s)}.

\begin{table}
\caption[A]{Theoretical ISB corrections, $\delta^{\text{(NCCI)}}_{\text{C}}$, (in \%)
adopted in this work. For the sake of comparison the table contains also the NSM results, $\delta^{\text{(NSM)}}_{\text{C}}$,
of Ref.~\cite{(Sev08s)}.
} \label{tab01}
\begin{tabular}{ccccccccc}
\hline\hline
 A    & $\,$ &    {~~~$\delta^{\text{(NCCI)}}_{\text{C}}$~~~} &   $\delta^{\text{(NSM)}}_{\text{C}}$ & $\quad\quad$ &   A   &   $\,$ &
 {~~~$\delta^{\text{(NCCI)}}_{\text{C}}$~~~}   &   $\delta^{\text{(NSM)}}_{\text{C}}$ \\
\hline
  17  &  &      0.166(17)   &      0.585(27)     & &  37  & &      0.907(91)     &    0.734(61)   \\
  19  &  &      0.339(34)   &      0.415(39)     & &  39  & &      0.318(32)     &    0.855(81)   \\
  21  &  &      0.300(30)   &      0.348(27)     & &  41  & &      0.426(43)     &    0.821(63)   \\
  23  &  &      0.316(32)   &      0.293(22)     & &  43  & &      0.690(69)     &    0.50(10)    \\
  25  &  &      0.413(41)   &      0.461(47)     & &  45  & &      0.589(59)     &    0.87(12)    \\
  27  &  &      0.439(44)   &      0.312(34)     & &  47  & &      0.673(67)     &                \\
  29  &  &      0.520(52)   &      0.976(53)     & &  49  & &      0.646(65)     &                \\
  31  &  &      0.585(59)   &      0.715(36)     & &  51  & &      0.714(71)     &                \\
  33  &  &      0.705(71)   &      0.865(59)     & &  53  & &      0.898(90)     &                \\
  35  &  &      0.366(37)   &      0.493(46)     & &  55  & &      0.620(62)     &                \\
\hline\hline
\end{tabular}
\end{table}

In summary, we have presented a systematic study of the GS$\rightarrow$GS GT and Fermi MEs in $T$=1/2 mirror
nuclei using, for the first time, the NCCI approach based on the isospin and angular momentum projected MR-DFT formalism.
The framework is universal and can be applied to any nucleus irrespectively on its mass and the proton and neutron
number parities. It can be also improved and optimized in many different ways, in particular, concerning the tensor
force which is known to have an impact on the shell structure~\cite{(Ots01cs),(Zal08s)} and
$\beta$-decay~\cite{(Min13)}.

In the present implementation with the SV or SV$_{\text{SO}}$ EDFs the calculated GT MEs systematically overestimate
experimental data for the free neutron strength of the axial current. The level of disagreement is
found to be very similar to the one obtained using large scale shell model in spite of the fundamental differences
between the two approaches in handling the core and the core polarization effects, the correlations or the basis truncation.
It strongly suggests, that the mechanism of in-medium renormalization of the axial strength may indeed be related to
the two-body currents. 
This conjecture requires further
studies.

Eventually, we have also calculated the ISB corrections to the Fermi decay branch in the $T$=1/2 mirror nuclei. The
corrections  turn out to be in a very good agreement with the NSM calculations.

This work was supported in part by the Polish National Science Centre (NCN) under Contracts 2012/07/B/ST2/03907
and 2014/15/N/ST2/03454. The CSC$-$IT Center for Science Ltd, Finland, is acknowledged for the allocation of computational
resources.


%

\end{document}